\documentclass[12pt]{amsart}

\usepackage{amssymb}

\def\conv{{\bf conv}}
\def\Supp{{\bf Supp}}
\def\mindef{{\bf Mindef}}

\title{Hypergraph partitions}
\author{A. S. Mishchenko$^1$, V. Manuilov$^2$, Chao You$^3$, Han Yang$^4$}

\date{}

\address{$^1$Moscow State University, Leninskie Gory 1, Moscow, 119991, Russia}

\email{asmish-prof@yandex.ru}

\address{$^2$Moscow State University, Leninskie Gory 1, Moscow, 119991, Russia}

\email{manuilov@mech.math.msu.su}

\address{$^3$Harbin Institute of Technology, Harbin 150001, Heilongjiang, China.}

\email{youchao@hit.edu.cn}

\address{$^4$Harbin Institute of Technology, Harbin 150001, Heilongjiang, China.}

\email{yanghanhuoying@live.com}

\begin{document}
\maketitle

\begin{abstract}
We suggest a reduction of the combinatorial problem of hypergraph partitioning to a continuous optimization problem. 

\end{abstract}

\section*{}

This paper is based on the papers by S. Schlag et al
\cite{schlag-2015}, Liu et al \cite{Liu-2011}, presented to us by HIT student Han Yang and discussed in October 2018 in Harbin.

Also we are aware about other papers on this topic, e.g. the survey by D. A. Papa and I. L. Markov \cite{Papa-2007}.

\section*{The work by S. Schlaget al,
``$k$-way Hypergraph Partitioning via $n$-Level Recursive Bisection''}
In the paper \cite{schlag-2015}, a multilevel algorithm for multigraph partitioning that contracts the vertices one at a time is developed. The running time is reduced by up to two-orders of magnitude compared to a naive $n$-level algorithm that would be adequate for ordinary graph partitioning. The overall performance is even better than the widely used hMetis hypergraph partitioner that uses a classical multilevel algorithm with few levels.
Considerably larger improvements are observed for some instance classes like social networks, for bipartitioning, and for partitions with an allowed imbalance of 10\%.
The algorithm presented in this work forms the basis of the hypergraph partitioning framework KaHyPar (Karlsruhe Hypergraph Partitioning).

\section{Introduction}

A hypergraph is a generalization of a graph, where each (hyper)edge can connect more than two vertices. The $k$-way partitioning problem for a hypergraph generalizes the well-known problem of graph partitioning:

How to divide the set of vertices into $k$ disjoint parts with sizes not exceeding $1+\varepsilon$ of the average block size, while the cost function, i.e. the sum of wieghts of all hyperedges that connect different parts should be minimized.

It is known that using hyperedges makes the partition problem more difficult \cite{curino-2010}, \cite{heintz-2010}.

Hypergraph partitioning (HGP) has a lot of applications. The two important areas of applications are VLSI circuit design and scientific calculations (e.g. speeding up sparse matrix-vector multiplications) \cite{Papa-2007}.
While the first one provides an example, where minor optimization can give sufficient effect, in the second one, modelling based on hypergraphs is more flexible than that based on graphs
\cite{heintz-2010}, \cite{catalyurek-1999}, \cite{hendrickson-2000}, \cite{klamt-2009}.

As the hypergraph partitioning is an NP-hard problem \cite{lengauer-1990} and as it is NP-hard even to find a good approximate solution for graphs \cite{bui-1992}, heuristic algorithms are usually used. The most often used heuristic algorithm is the multilevel paradigm, which 
consists of three
phases: In the coarsening phase, the hypergraph is
recursively coarsened to obtain a hierarchy of smaller
hypergraphs that reflect the basic structure of the input.
After applying an initial partitioning algorithm to the
smallest hypergraph in the second phase, coarsening is
undone and, at each level, a local search method is used
to improve the partition induced by the coarser level.

\section{Combinatorial formulation of the problem}

Since the problem of hypergraph partitioning is formulated approximately, we suggest to replace the original problem by its approximation from the very beginning.

So, we start with a hypergraph $\Gamma$, consisting of a finite number of vertices $V=V(\Gamma)$ and a finite number of hyperedges. Each hyperedge $e\in E(\Gamma)$ is given by its ends, which are connected by this hyperedge, i.e. by a finite subset $End(e)\subset V(\Gamma), \quad \#(End(e))<\infty.$ In particular, among the hyperedges, there may be simplest edges, that connect only two vertices, i.e. such hyperedges $e$ that $\#(End(e))=2$.

The $k$-partitioning problem for a hypergraph $\Gamma$ can be formulated as follows: to find subsets
$\Gamma_{1}\subset \Gamma,$ $\Gamma_{2}\subset \Gamma,$ \dots $\Gamma_{k}\subset \Gamma,$ such that:
\begin{enumerate}
\item they are disjoint;
\item $\#\Gamma_{i}\approx \frac{1}{k}\sum_{i}\#\Gamma_{i}$ up to $\varepsilon$;
\item  the number of hyperedges that connect vertices from different subsets $\Gamma_{i}, \Gamma_{j}$ is minimal.
\end{enumerate}

\section{Reduction of the combinatorial problem to a continuous problem}

Consider first the simplest case of the combinatoriaal problem, when the hypergraph $\Gamma$ is a classical one-dimensional graph, i.e. all edges are one-dimensional.

Consider then the simplex $\Delta$ generated by vertices $V(\Gamma)$. Everything happens on the space $\Delta$. Each vertex is identified with the delta-function on $\Delta$ with this vertex being its support. Therefore, we may replace the set of vertices by the space of functions $C(\Delta)$. If the graph $\Delta$ is partitioned into two parts, $\Delta=\Delta_{1}\sqcup\Delta_{2}$ then, instead of these parts, $\Delta_{1}$, $\Delta_{2}$, we consider two functions,
$f_0,f_1\in C(\Delta)$, such that 
$$
f_{1}\geq 0\quad\mbox{ and}\quad f_{2}\geq 0;
$$ 
$$
\Supp f_{i} = \conv(\Delta_{i}).
$$
The requirement $\Delta_{1}\cap\Delta_{2}=\emptyset$ can be replaced by the requirement
$$f_{1}(x)\cdot f_{2}(x) \equiv 0, \quad x\in\Delta,$$
or, approximately, by
$$
\max_{x\in\Delta}|f_{1}(x)\cdot f_{2}(x)|\leq  \varepsilon.
$$

The size of a part is measured by the integral
$$
\int\limits_{\Delta} f_{i}(x)dx,
$$
which should be approximately equal to the average part size, i.e. 
$$
\left|\int\limits_{\Delta} f_{i}(x)dx-\frac{1}{2}\#(\Delta)\right|\leq\varepsilon.
$$

Each edge of the graph $\Gamma$ can be described as a function $g(x,y)$ on the Cartesian product $\Delta\times\Delta$. This function should approximate the edge $(a,b)\in V(\Gamma)\times V(\Gamma))$ by using the support of the function $g(x,y)$.
Then the number of edges connecting the two parts can be written as
$$
F(x,y)=f_{i}(x)f_{2}(y)g(x,y).
$$

Therefore, the problem is reduced to minimizing the integral
$$
\mindef\left(\int\limits_{(x,y)\in\Delta\times\Delta}F(x,y)dxdy\right).
$$

Summing up, the problem reduces to the following one:
Find functions $f_{i}(x)$ and $f_{2}(x)$, $x\in\Delta$, satisfying the conditions:
\begin{itemize}
\item  The condition of disjointness: $$\max_{x\in\Delta}|f_{1}(x)\cdot f_{2}(x)|\leq  \varepsilon.$$
\item The condition of almost equal sizes: $$\left|\int\limits_{\Delta} f_{i}(x)dx-\frac{1}{2}\#(\Delta)\right|\leq\varepsilon.$$
\item minimizing the integral
$$
\mindef\left(\int\limits_{(x,y)\in\Delta\times\Delta}F(x,y)dxdy\right)=
\mindef\left(\int\limits_{(x,y)\in\Delta\times\Delta}f_{i}(x)f_{2}(y)g(x,y)dxdy\right).
$$
\end{itemize}

The formulation of the problem can be naturally transferred to the case of hypergraphs, where edges are replaced by hyperedges, and the number of parts can be greater than two.

\section{Solution of the analytical problem}

Note that the condition of disjointness is of different nature than the two other conditions, namely, it should be checked at each point of $V$ separately, while the two other conditions are integrals. We may replace the disjointness condition by a weaker one:
$$\int\limits_{\Delta}f_1(x)f_2(x)\,dx\leq\varepsilon\#(\Delta).$$

In this way we may get a few points, where both $f_1$ and $f_2$ are not small, but the number of such points cannot be too great.

Let us also replace $f_2$ by $1-f_1$, and the problem reduces to that of finding a function $f\in C(\Delta)$ such that $f$ satisfies the two conditions:
\begin{itemize}
\item
$$\int\limits_{\Delta}f(x)(1-f(x))\,dx\leq\varepsilon\#(\Delta);$$
\item
$$\left|\int\limits_{\Delta} f(x)dx-\frac{1}{2}\#(\Delta)\right|\leq\varepsilon,$$
\end{itemize}
and minimizes the integral
$$
\mindef\left(\int\limits_{(x,y)\in\Delta\times\Delta}f(x)(1-f(y))g(x,y)dxdy\right).
$$
This can be written in a matrix form. Let $G$ denote the matrix of $g(x,y)$, $a=(1,1,\ldots,1)$ the vector with all coordinates equal to 1. To simplify the notation, let also $\frac{1}{2}\#(\Delta)=C$ Then the above conditions are:
\begin{itemize}
\item
$$\langle f,(a-f)\rangle\leq 2\varepsilon C;$$
\item
$$|\langle a,f\rangle-C|\leq\varepsilon;$$
\item
$$
\langle f,G(a-f)\rangle\to \min.
$$
\end{itemize}
This can be solved by using the Lagrange multipliers method. We have to minimize the functional
$$f\mapsto \langle f,G(a-f)\rangle-\lambda(\langle f,(a-f)\rangle-2\varepsilon C)-\mu((\langle a,f\rangle-C)^2-\varepsilon^2).$$
The critical points of this functional satisfy
$$
\langle df,G(a-f)\rangle-\langle f,Gdf\rangle-\lambda\langle df,(a-f)\rangle+\lambda\langle f,df\rangle-2\mu(\langle a,f\rangle-C)\langle a,df\rangle=0
$$
for any $df$.

When the matrix $G$ is symmetric (which is natural for adjacency matrices), we may rewrite this as
$$
\langle df,G(a-f)-Gf-\lambda(a-f)+\lambda f-2\mu(\langle a,f\rangle-C)a\rangle=0,
$$
hence the critical points of the functional should satisfy
$$
(G-\lambda)(a-2f)=2\mu(\langle a,f\rangle-C)a,
$$
together with
$$\langle f,(a-f)\rangle\leq 2\varepsilon C$$
and
$$|\langle a,f\rangle-C|\leq\varepsilon.$$

\end{document}